
\font\bigbf=cmb10 scaled 1200
\nopagenumbers
\headline={\ifnum\pageno=1 \hfil NUB 3053/92 \else \hss -- \folio\ --\hss \fi}
\vglue 1in
\centerline{\bigbf \hskip-24pt Wightman Functions in QCD}
\vskip .5in
\centerline{M. H. Friedman, Y. Srivastava and A. Widom}
\vskip .5in
\centerline{\it Physics Department}
\centerline{\it Northeastern University}
\centerline{\it Boston, MA 02115, USA}
\centerline{\it and}
\centerline{\it Dipartimento di Fisica \& INFN}
\centerline{\it  Universita di Perugia}
\centerline{\it Perugia, Italy}
\vskip 1in
\centerline{\it ABSTRACT}

\par The constraint imposed by Gauss' law is used to show that the
matrix elements of n-point Wightman Functions of gluon field and
quark current operators at different space time points vanish
when taken between physical states.
\vfil\eject
\noindent {\bf 1. Introduction.}

\par
Christ and Lee[1,2] have warned against the common practice of
starting from a formal path integral approach prior to choosing a
gauge. In fact, for QCD they chose the (time)-axial (or temporal)
gauge[3,4,5] which has a well defined Hamiltonian and then derived the
Feynman rules for different gauges by appropriate coordinate
transformations. In the temporal gauge, their construction of the path
integral generating functional requires for consistency that it
operate only between color singlet states.
\par
For our purposes it will be convenient to work in the canonical
formalism[5]. In section 2 we point out that just
implementing the constraints, due to Gauss' law, on physical states in
the temporal gauge leads us to a pleasing result: all transition
matrix elements of the color electric and magnetic fields as well as
of color carrying currents between physical states vanish. This
forecloses the possibility of any physical state with non-zero color
appearing. In section 3 we show that two point Wightman
functions of quark current and gluon fields in physical states vanish
when the two points do not coincide, and in section 4 we extend the
results to n-point functions.
\vskip .3in
\noindent {\bf 2. The Physical States.}

\par
In the temporal gauge, $A_0^a=0$, the color electric field
$$
E^{a}_{i}({\bf x},t)=-\partial A^{a}_{i}({\bf x},t)/\partial t.
\eqno(1)$$
\par
We will state the canonical commutation relations a little more
carefully than is customary, so as to avoid inconsistencies. To this
end let us consider the state vector $\Psi$ which is a member of an
Hilbert space spanned by a complete set of normalizable functions of
the vector potentials $A^{a}_{i}$, obeying appropriate boundary
conditions. Let $\cal F$ be an operator such that ${\cal F}\Psi$ is
also a member of this Hilbert space. Then the canonical equal time
commutation relations are given by
$$
[E^{a}_{i}({\bf x},t),{\cal F}(t)] = i{\delta {\cal F}(t) \over
{\delta A^a_i({\bf x},t)}}.
\eqno(2)$$
\par
The Gauss' law operator $G^a({\bf x},t)$
$$
G^{a}({\bf x},t) = \partial _{i}E^{a}_{i}({\bf x},t) -
gf^{abc}A^{b}_{i}({\bf x},t)E^{c}_{i}({\bf x},t) - J^{a}_{0}({\bf
x},t),
\eqno(3)$$
constraint on physical states $\vert \gamma \rangle$ reads[1,3,4,5,6,7,8,9,10]
$$
G^{a}({\bf x},t) \, \vert \gamma \rangle = 0.
\eqno(4)$$
Unlike the QED case, the Gauss operator $G^a$ generates rotations in
color space and as such rotates the color electric $E_i^a$ and
magnetic $B_i^a$ fields as well. For example,
$$
[G^{a}({\bf x},t),F^{b}_{\mu\nu}({\bf x}',t)] =
igf^{abc}F^{c}_{\mu\nu}({\bf x},t)\delta ^{3}({\bf x}-{\bf x}').
\eqno(5)$$
where $F^a_{\mu\nu}$ is the color electromagnetic field tensor.
\par
Taking the matrix element of Eq.(5) between arbitrary physical states
$\gamma$ and $\gamma '$ and using Eq.(4) we find that
$$
\langle \gamma '| F^{a}_{\mu\nu}({\bf x},t) | \gamma  \rangle = 0.
\eqno(6)$$
\par
An identical answer ensues for the color current density $J_{\mu}^a$,
viz.,
$$
\langle \gamma '| J^{a}_{\mu }({\bf x},t) | \gamma  \rangle = 0.
\eqno(8)$$
Vanishing of all of the above can be satisfied only for physical
states which are color singlets[11]. The equivalence of temporal and
Coulomb gauges in QCD[1,2], would then imply that an identical result holds
there as well. Note that it is only for gauge covariant
quantities that one can meaningfully speak about their value. If they
become null identically in one gauge they remain so in any other
gauge. In contrast, we do not have a similar argument regarding non
gauge covariant quantities such as the vector potential. We further
note that we have assumed there is no symmetry breaking, so that a
similar argument cannot be applied to $SU(2)\times U(1)$.
\vskip .3in
\noindent {\bf 3. Two-Point Wightman Functions.}

\par
Having deduced that the physical matrix elements of non-color singlet
operators vanish, we next turn to constraining the matrix
elements of color-singlet operators which are composites of other
colored operators. It is not difficult to show that these are devoid
of absorptive parts. By way of illustration consider the commutator of
the product of YM fields $F^b_{\mu\nu}$ and $F^c_{\xi \sigma}$, with the Gauss
law operator $G^a$. Using Eq.(5) and  the fact that the
$G^a(x)$ are independent of $x^0$, we find
$$
f^{abc} \left[ G^a(x),F^b_{\mu\nu}(y)F^c_{\xi \sigma}(z)\right] =
3igF^a_{\mu\nu}(y)F^a_{\xi \sigma}(z)\left\{ \delta^3({\bf x}-{\bf y})
-\delta^3({\bf x}-{\bf z})\right\}.
\eqno(9)$$
Taking Eq.(9) between arbitrary physical states $\gamma$ and $\gamma
'$ and using Eq.(4) yields
$$\left< \gamma '\vert F^a_{\mu\nu}(y)F^a_{\xi \sigma}(z) \vert \gamma
\right> = 0 \eqno(10)$$
when {\bf y} $\not=$ {\bf z}.
Employing the Lorentz invariance of the theory we then find that it is
true for all $y^\mu\not= z^\mu$. A similar argument holds for the
quark current operators either by themselves or in combination with
$F^a_{\mu\nu}$.
\vskip .3in
\noindent {\bf 4. N-point Functions.}

We now construct operators $V^a(x)$ which are composites of current
density and gluon fields at the same space time point and transform as
vectors under the gauge group (for convenience). (In general they will
also carry Lorentz indices as well, but which we now suppress). Let us then
consider the following matrix elements in physical states of the
commutators
$$
<\alpha| [G^a(x), V^{b_1}(y_1)V^{b_2}(y_2)\cdots V^{b_n}(y_n)] |\beta>
=
$$
$$
ig<\alpha| [f^{ab_1c}V^c(y_1)V^{b_2}(y_2)\cdots
V^{b_n}(y_n)\delta^3({\bf x}-{\bf y}_1)
$$
$$
+ f^{ab_2c}V^{b_1}(y_1)V^c(y_2)\cdots V^{b_n}(y_n)\delta^3
({\bf x}-{\bf y}_2)
$$
$$+ \cdots +  f^{ab_nc}V^{b_1}(y_1)V^{b_2}(y_2)\cdots V^c(y_n)
\delta^3({\bf x}-{\bf y}_n)] |\beta> = 0 \eqno(11)
$$
using Eq.(4). Thus, if ${\bf y}_i \ne {\bf y}_j$ for all $i \ne j$
then each of the coefficients of the $\delta^3({\bf x}-{\bf y}_i)$
must vanish. As before, the $y_i^0$ are arbitrary. Hence, once again,
appealing to the Lorentz invariance of the theory, we find that these
coefficients must vanish for space like, time like and light like
separations of all the $y_i$.
\par
Furthermore, there is only one $f^{abc}$ for a given $a$ and $b$.
Thus, $f^{abc} V^c(y)$ is a single term and the $n$-point functions
must vanish unless two or more of the $y_i$ are equal. In this latter
case we may attempt to extract the singlet content at such points by
contracting with $f^{abc}$, $d^{abc}$ and $\delta^{ab}$. If after all
such contractions there remain terms with non zero color content at
one or more points $y_i$ then they become matrix elements
belonging to a smaller $n$ and hence they must also vanish. Matrix
elements which have only singlet operators at different space time
points need not vanish.
\vskip .3in
\noindent {\bf 5. Conclusions.}

\par
The physical states of QCD are ones which satisfy Gauss' law and hence
have zero color. Furthermore, Wightman functions in physical states,
which involve color non-singlet operators at different space time
points, vanish identically.
\vfill\eject
\noindent {\bf Acknowledgements.}

\par
This work was supported in part by INFN in Italy and by DOE in the
United States.
\vskip .4in
\centerline{\bf REFERENCES}
\par \noindent
[1] T. D. Lee, "Particle Physics and Introduction to Field Theory",
Harwood Academic Publishers, NY (1988), Chapters 18 and 19.
\par\noindent
[2] N.M. Christ and T.D. Lee, Phys. Rev. {\bf D22}, 939 (1980).
\par \noindent
[3] J. Schwinger, Phys. Rev. {\bf 127}, 324 (1962); {\bf
130}, 406 (1963).
\par\noindent
[4] R. P. Feynman, Acta Physica Polonica {\bf 24} 697
 (1963).
\par\noindent
[5] H. Weyl, "The theory of groups and Quantum Mechanics", Dover
Publications, New York, NY, first published in 1950. (It is a
republication of the English translation of {\it Gruppentheorie und
Quantenmechanik} originally published in 1931).
\par\noindent
[6] B. Sakita, "Quantum Theory of Many-Variable Systems
and Fields", World Scientific Co. Pvt. Ltd (1985), Singapore.
\par\noindent
[7] J. L. Gervais and B. Sakita, Phys. Rev. {\bf D18},
 453 (1978).
\par\noindent
[8] J. Kogut and L. Susskind, Phys. Rev. {\bf D9}, 3501
 (1974); {\bf D11}, 395 (1975).
 \par\noindent
[9] S. Mandelstam, Annals of Phys. {\bf 19}, 1 (1962).
\par\noindent
[10] A. M. Polyakov, "Gauge Fields and Strings", Harwood Academic
Publishers, NY (1987)
\par\noindent
[11] A similar argument was presented by K. Cahill in a preprint,
Indiana University, IUHET 34 (1978), (unpublished).
\par\noindent

\bye